\documentclass[
 reprint,
superscriptaddress,
 amsmath,amssymb,
 aps, 
 prl, 
floatfix,
]{revtex4-2}

\usepackage[dvipsnames]{xcolor}
\usepackage{soul}

\usepackage{physics}

\usepackage{dcolumn}
\usepackage{bm}

\usepackage[utf8]{inputenc}
\usepackage{amsmath}
\usepackage{graphicx}
\usepackage{svg}
\usepackage{hyperref}

\newcommand{\Vs}{V_\text{s}}
\newcommand{\kt}{k_\text{B}T}
\newcommand{\eps}{\varepsilon}
\newcommand{\epsls}{\varepsilon_\text{LS}}

\newcommand{\ilm}{Univ Lyon, Univ Claude Bernard Lyon 1, CNRS, Institut Lumi\`ere Mati\`ere, F-69622, VILLEURBANNE, France}

\begin{document}

\preprint{APS/123-QED}

\title{Giant slip length at a supercooled liquid-solid interface}

\author{Suzanne Lafon}
 \email{suzanne.lafon@universite-paris-saclay.fr}
 \affiliation{Paris-Saclay University, CNRS, Solid State Physics Laboratory, 91405 Orsay, France}
\author{Alexis Chennevière}
 \affiliation{CEA Saclay, Léon Brillouin Laboratory, 91191 Gif-sur-Yvette, France}
\author{Frédéric Restagno}
 \affiliation{Paris-Saclay University, CNRS, Solid State Physics Laboratory, 91405 Orsay, France}
\author{Samy Merabia}
 \affiliation{\ilm}
\author{Laurent Joly}
 \affiliation{\ilm}
 %\affiliation{\iuf}

\date{\today}

\begin{abstract}
The effect of temperature on friction and slip at the liquid-solid interface has attracted attention over the last twenty years, both numerically and experimentally. However, the role of temperature on slip close to the glass transition has been less explored. Here, we use molecular dynamics to simulate a bi-disperse atomic fluid, which can remain liquid below its melting point (supercooled state), to study the effect of temperature on friction and slip length between the liquid and a smooth apolar wall, in a broad range of temperatures. At high temperatures, an Arrhenius law fits well the temperature dependence of viscosity, friction and slip length. In contrast, when the fluid is supercooled, the viscosity becomes super-Arrhenian, while interfacial friction can remain Arrhenian or even drastically decrease when lowering the temperature, resulting in a massive increase of the slip length. We rationalize the observed superlubricity by the surface crystallization of the fluid, and the incommensurability between the structures of the fluid interfacial layer and of the wall. This study calls for experimental investigation of the slip length of supercooled liquids on low surface energy solids.

\end{abstract}

\maketitle

\section{Introduction}

In 1823, Navier \cite{navier} postulated the existence of a velocity jump at the liquid-solid interface and proposed a linear relation between the interfacial stress and the velocity jump: $\tau_{\text{LS}} = \lambda\, \Vs$, where $\lambda$ is the interfacial friction coefficient and $\Vs$ is the velocity jump, also called the slip velocity. Because of stress continuity, this is equal to the bulk shear stress $\tau_\text{bulk} = \eta\, \dot{\gamma}$, with $\eta$ the fluid viscosity and $\dot{\gamma}$ the shear rate for sufficiently low shear rates \cite{priezjev_molecular_2004}. A more classical way of characterizing slip at the solid-liquid interface is to introduce the slip length $b$, which is the length at which the velocity profile of the liquid linearly extrapolates to the velocity of the wall \cite{neto_boundary_2005}, leading to:
\begin{equation}
    b = \frac{V_{s}}{\dot{\gamma}} = \frac{\eta}{\lambda}. 
\end{equation}
The slip length $b$ is thus dependent on the liquid-surface interaction through the friction coefficient $\lambda$. Since both the viscosity $\eta$ and the friction coefficient $\lambda$ depend on temperature, so does the slip length $b$. However, $\eta$ and $\lambda$ are both decreasing functions of temperature, thus the resulting effect of temperature on the slip length is not trivial.

In the literature, various behaviors of $b(T)$ have been observed. Experimentally, \citet{baumchenSlidingFluidsDewetting2010} reported a decreasing $b(T)$ for PS thin films, while \citet{drdaStickSlipTransitionPolymer1995} observed an almost constant slip length for PE melts. Using molecular dynamics (MD) simulations, \citet{servantieTemperatureDependenceSlip2008} reported non-monotonic variations of the slip length of a Lennard-Jones (LJ) polymer, and \citet{herreroFastIncreaseNanofluidic2020b} measured a decreasing $b(T)$ for water and methanol on different types of surfaces. In addition, \citet{andrienkoBoundarySlipResult2003} predicted a jump of the slip length at low temperatures because of prewetting transition at the interface between a binary mixture and a solid wall.

Different models have been proposed in order to rationalize the temperature dependence of the viscosity $\eta$, the friction coefficient $\lambda$ and the slip length $b$. A simple description is Eyring's theory, which assumes that flow is an activated process: in order to jump from one position to a neighbouring one, a given molecule has to overcome an energy barrier $E_{\text{a}}$. Although it has been shown that the real microscopic dynamics is not a barrier hoping mechanism \cite{hansen2013theory,rizkMicroscopicOriginsViscosity2022}, Eyring's theory is still useful to compare the general temperature dependency of $\eta$ and $\lambda$ in ordinary liquids.
%In particular, although its physical relevance in liquids is still under investigation \cite{rizk2022microscopic}, Eyring theory (also called molecular kinetic theory) assumes that flow is an activated process: in order to jump from one position to a neighbouring one, a given molecule has to overcome an energy barrier $E_{\text{a}}$. This
It can be applied both to the bulk flow, leading to an Arrhenian viscosity $\eta \propto \text{exp} \left\{E_{\text{a,viscous}}/(\kt)\right\}$ \cite{pelzActivationEnergyWall2021}, and to the flow near the wall, leading to an Arrhenian friction coefficient $\lambda \propto \text{exp} \left\{E_{\text{a,friction}}/(\kt)\right\}$ \cite{blakeSlipLiquidSolid1990,jacobsGrowthHolesLiquid1998}. Therefore, the slip length also follows an Arrhenius law \cite{lichterLiquidSlipNanoscale2007,wangSlipBoundaryConditions2011,wangUniversalMolecularkineticScaling2019,guoTemperatureDependenceVelocity2005,pelzActivationEnergyWall2021}, which can be expressed as:
\begin{equation}
    b \propto \text{exp} \left(\frac{E_{\text{a,viscous}}-E_{\text{a,friction}}}{\kt}\right), 
    \label{Eyring_b(t)}
\end{equation}
and one cannot know \textit{a priori} its variation with temperature. Recently, Hénot \textit{et al.} \cite{henotTemperatureControlledSlipPolymer2018} have used this formalism to discuss the effect of temperature on the slip length of PDMS melts measured with a velocimetry technique. Equation~\eqref{Eyring_b(t)} fits well their data, and depending on the surface, $E_{\text{a,friction}}$ was either larger than or equal to $E_{\text{a,viscous}}$, implying that the slip length was increasing or constant with temperature, respectively.

However, this Arrhenius picture is not always accurate, especially for supercooled liquids. Indeed, near the glass transition, the viscosity increases much more sharply than an Arrhenian dependency. To account for this quick increase, other models have been proposed, among which the Vogel-Fulcher-Tammann (VFT) law \cite{vogelViscosity1921,fulcherViscosity1925,tammannViscosity1926}, which states that $\eta \propto \text{exp} \left\{A/(T-T_\text{VFT})\right\}$, with $T_\text{VFT}$ a reference temperature at which the viscosity diverges. This law is widely used to describe the temperature dependence of the viscosity close to the glass transition, and it has also been used to describe the temperature dependence of the friction coefficient \cite{herreroFastIncreaseNanofluidic2020b}.

In this article, we present MD computations of the slip length of a model bi-disperse LJ liquid in a wide range of temperatures. Far from the glass transition temperature, we show that the slip length is Arrhenian with an effective activation energy controlled by the strength of the liquid-solid (L-S) interaction: $b(T)$ decreases with temperature for weak L-S coupling while it increases with temperature for strong interaction with the wall. At lower temperatures, the slip length may increase by orders of magnitude as a result of the super-Arrhenian behavior of the viscosity, and the Arrhenian or even sub-Arrhenian behavior of the friction coefficient. In particular, for weakly interacting surfaces, the first liquid layers become structured and the incommensurability between the fluid local structure and the wall lattice results in a strong reduction of the L-S friction, and thus giant values of the slip length.

\begin{figure}
	\centering
	\includegraphics[scale=1]{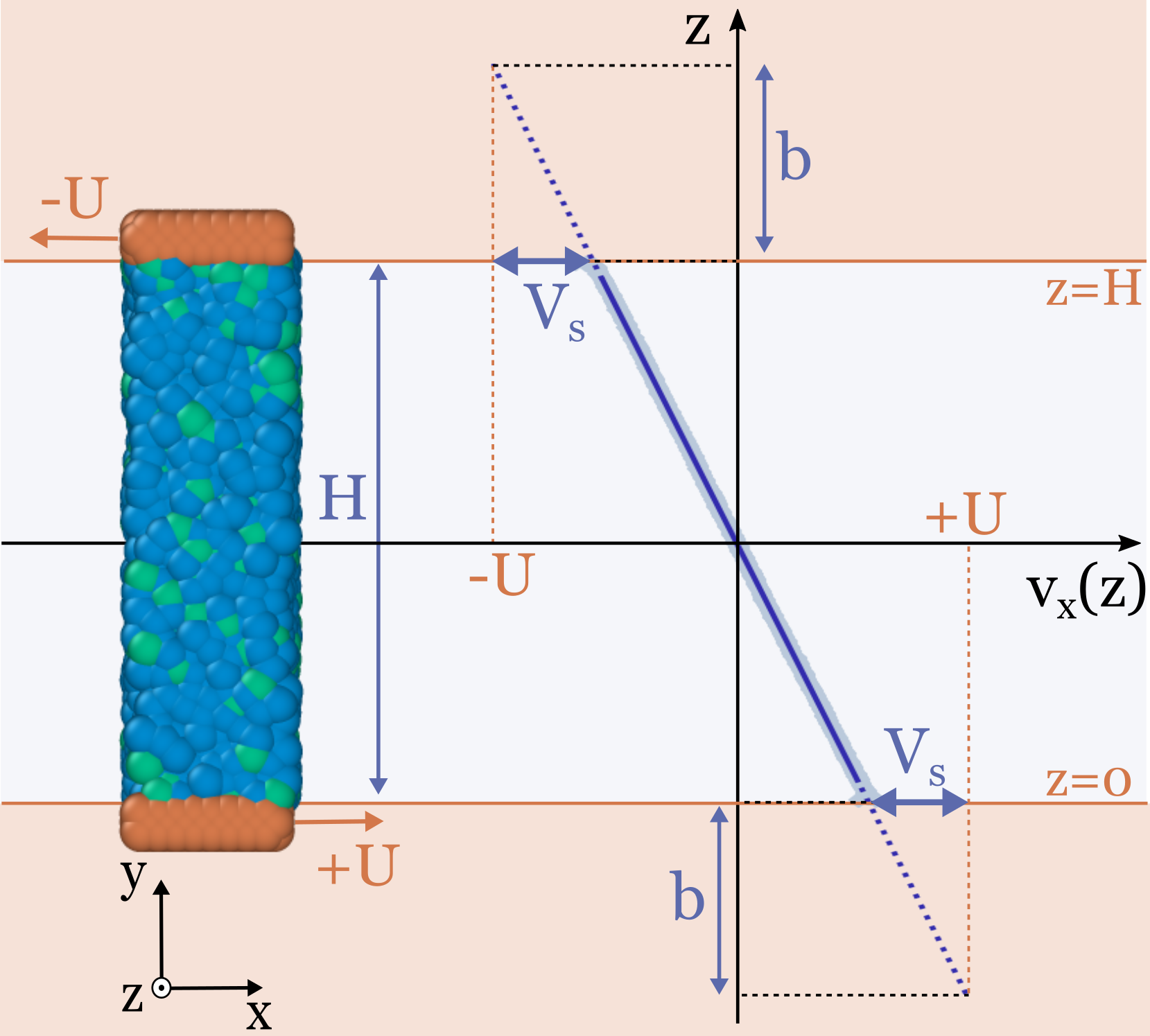}
	\caption{Schematic of the simulated system and corresponding flow profile $v_{x}(z)$ of the liquid. The bottom (top) wall has a velocity $+U$ ($-U$) along the $x$ direction. The distance between the walls is denoted $H$. The slip length $b$ is defined as the length at which the velocity profile extrapolates to the velocity of the wall. The slip velocity $V_{s}$ is the velocity difference between the velocity of the liquid at the wall and the velocity of the wall.}
	\label{fig_schema}
\end{figure}

\section{Methods}

We perform MD simulations using the LAMMPS package \cite{LAMMPS}. We simulate a shear flow of a Kob-Andersen (KA) binary LJ liquid \cite{KobAndersen} sheared between two LJ walls (see Fig.~\ref{fig_schema}). We use the KA liquid described in \cite{pedersenPhaseDiagramKobAndersenType2018}, which is a mixture of two particle types $i= \text{A},\text{B}$ in a $80-20$ ratio. They interact through a LJ pair potential $V_{ij}(r) = 4 \eps_{ij} \left[ (\sigma_{ij}/r)^{12} - (\sigma_{ij}/r)^{6} \right]$, with $\eps_{\text{AA}}=\eps$ and $\sigma_{\text{AA}}=\sigma$ taken as references. 
In the following, all quantities are reported in LJ reduced units, using $\eps$, $\sigma$, and the particle mass $m$ as units of energy, distance, and mass, respectively, and taking $k_{\text{B}}=1$. All the atoms are supposed to have the same mass $m$, and we take $\sigma_{\text{BB}} = 0.88$, $\sigma_{\text{AB}} = 0.80$, $\eps_{\text{BB}} = 0.50$ and $\eps_{\text{AB}} = 1.50$. The potential is truncated and shifted to zero at $r_\text{c} = 2.5$. The wall is a crystallized face-centered cubic (FCC) lattice of type C particles with a lattice parameter $a = 1$, corresponding to a number density $\rho = 4.0$. The effect of the wall density on slip is discussed in the SM Fig.~S5.
The strength of the liquid-solid interaction potential $\eps_{\text{AC}} = \eps_{\text{BC}} \equiv \epsls$ is varied from $0.15$ to $1.00$ and we take $\sigma_{\text{AC}} = \sigma_{\text{BC}} = 1.00$ for all the simulations. 
%\hl{Following} \citet{pedersenPhaseDiagramKobAndersenType2018}, \hl{we take $T_{\text{m}} = 1.01$ and $T_{\text{g}} = 0.60$ as the melting and glass transition temperatures of our KA fluid, respectively.}
The wall dimensions are $L_{x} = L_{y} = 8.0$ with periodic boundary conditions in both $x$ and $y$ directions.

The temperature is imposed using a Nosé-Hoover thermostat with a damping time of $100$ time steps. When the liquid is sheared, the thermostat is coupled only to transverse velocities. The pressure is set to $10.0$ by using the top wall as a piston during a preliminary run. This is a standard choice of pressure for a Kob-Andersen liquid \cite{KobAndersen,pedersenPhaseDiagramKobAndersenType2018}, and we have checked that the pressure did not impact the slip length significantly (see Fig.~S4).
The top wall is then fixed at its equilibrium position. From this equilibrated system, we use two different procedures to measure $\eta$ and $\lambda$. In the first one, the walls are displaced along the $x$-axis at constant velocity $\pm U$. We record the velocity profile of the liquid $v_{x}(z)$ (Fig.~\ref{fig_schema}) and the stress exerted by the liquid on the walls $\tau_{LS}$. The viscosity $\eta$ is calculated with $\eta = \tau_{\text{LS}}/\dot{\gamma}$, with $\dot{\gamma}$ being the shear rate extracted from the velocity profile. We measure the hydrodynamic height $h$ using the Gibbs dividing plane (GDP) method described in \cite{herreroShearForceMeasurement2019}, see the supplemental material (SM), Fig.~S1. The friction coefficient $\lambda$ is then calculated by $\lambda = \tau_{\text{LS}}/\Vs$, with the slip velocity $\Vs = U - \dot{\gamma} h/2$. The shear velocity $U$ is varied between $0.001$ and $1.20$ and the values of $\eta$ and $\lambda$ are taken in the linear response regime (see the SM, Fig.~S2). The details of the procedure are described in the SM, Fig.~S3.

The second procedure consists in measuring both parameters at equilibrium  using Green-Kubo relations \cite{hansen2013theory,bocquet2013green}. For the viscosity one uses: 
\begin{equation}
    \eta = \frac{V}{k_{B}T} \frac{1}{5} \sum_{i} \lim\limits_{t \to +\infty} \int_{0}^{t} \langle\sigma_{i}(0) \sigma_{i}(\tau)\rangle \dd \tau , 
\end{equation}
where $V$ is the volume, $k_{\text{B}}$ is the Boltzmann constant (here taken equal to $1$), $T$ is the temperature of the system, and the  $\sigma_{i} = \sigma_{xy}, \sigma_{xz}, \sigma_{yz}, (\sigma_{xx}-\sigma_{yy})/2, (\sigma_{yy}-\sigma_{zz})/2$ are the traceless components of the stress tensor inside the liquid and are measured in an independent, fully periodic simulation. 
For the friction coefficient, one uses: 
\begin{equation}
    \lambda = \frac{S}{k_{B}T} \frac{1}{2} \sum_{j} \lim\limits_{t \to +\infty} \int_{0}^{t} \langle\sigma_{j}(0) \sigma_{j}(\tau)\rangle \dd \tau , 
\end{equation}
where $S=L_{\text{x}}L_{\text{y}}$ is the surface, and the $\sigma_{j} = \sigma_{\text{LS, top}}, \sigma_{\text{LS, bottom}}$ are the liquid-solid friction forces per unit surface along the $x$ direction at the two liquid-solid interfaces.

\section{High temperature regime ($T\geq 1.5$)}

\begin{figure}
	\centering
	\includegraphics[scale=1]{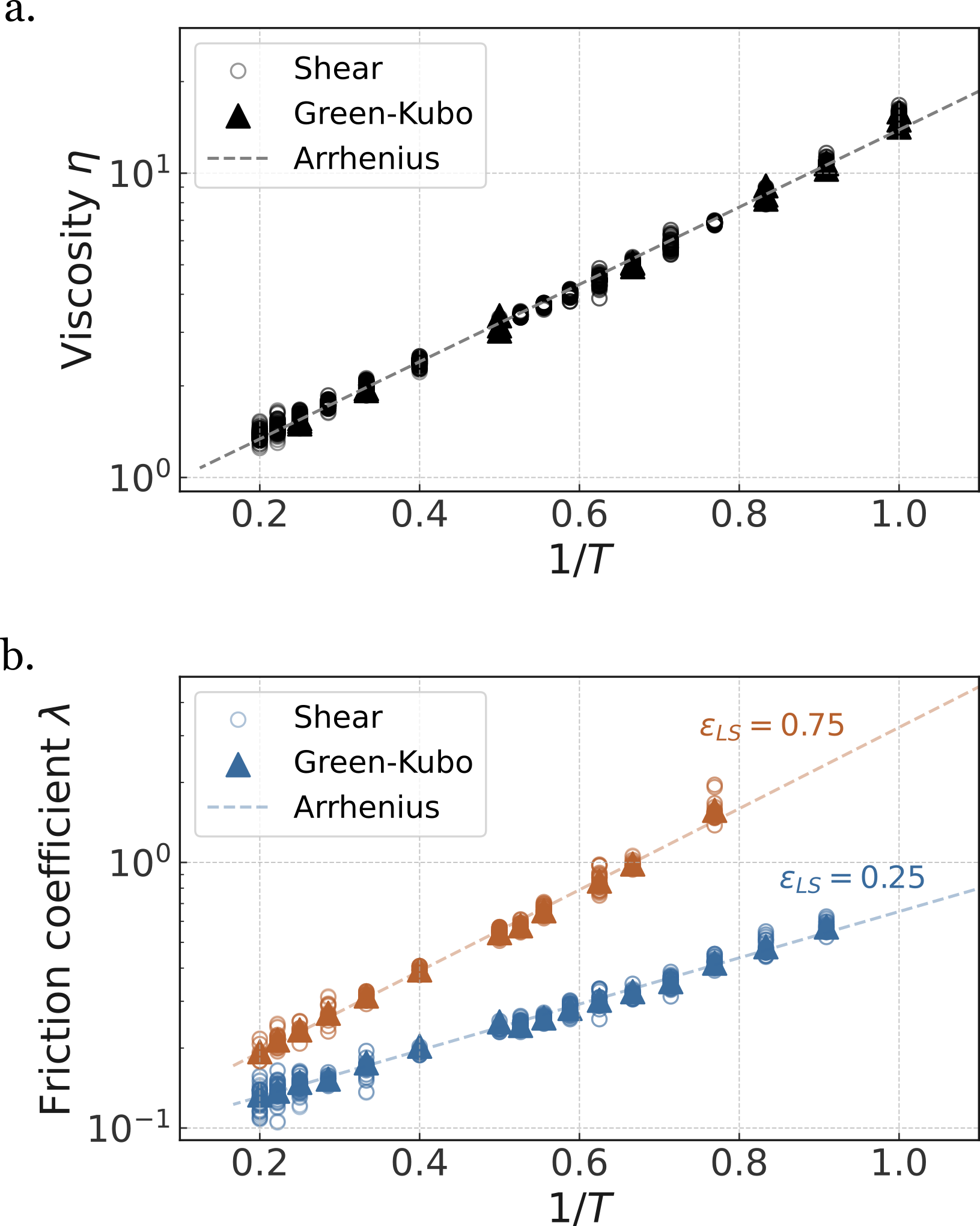}
	\caption{Viscosity $\eta$ and friction coefficient $\lambda$ as a function of temperature measured by shear (empty circles) and at equilibrium using Green-Kubo relations (filled triangles). Both methods give the same values. The friction coefficient measurement is illustrated here for two different values of the liquid-solid interaction energy $\epsls = 0.25$ (blue) and $\epsls = 0.75$ (orange). The dotted lines correspond to Arrhenius regressions.}
	\label{fig_GKshear}
\end{figure}

The results are shown in Fig.\ref{fig_GKshear}. At high temperatures, both the equilibrium and the non-equilibrium procedures give the same results for the viscosity and the friction coefficient. Both $\eta$ and $\lambda$ can be fitted by an Arrhenius law $\eta \propto \text{exp}(E_{\text{a,viscous}}/T)$ and $\lambda \propto \text{exp}(E_{\text{a,friction}}/T)$, with a friction activation energy $E_{\text{a,friction}}$ which depends on the strength of the liquid-solid interaction $\epsilon_{\text{LS}}$.

\begin{figure}
	\centering
	\includegraphics[scale=1]{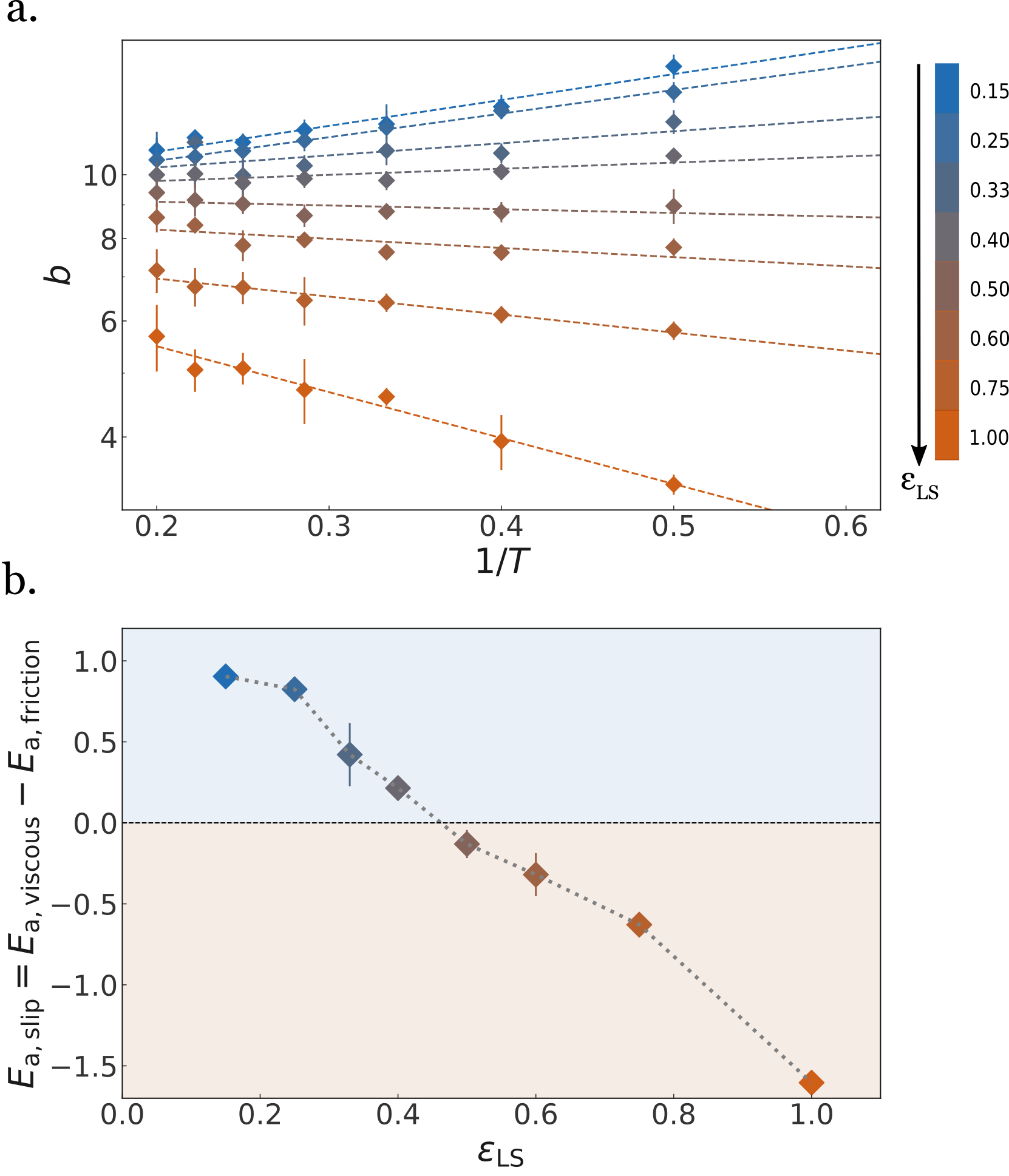}
	\caption{(a) Temperature dependence of the slip length for different liquid-wall interaction energies $\epsls$. The dotted lines correspond to Arrhenius-like regressions, from which we extract an activation energy for the slip length $E_{\text{a,slip}}$.
	%Because $b=\eta/\lambda$, the activation energy for slip arises as the difference between the activation energy for viscosity, and the activation energy for interfacial friction.
	(b) Activation energy of the slip length as a function of $\epsls$. The dotted line is a guide for the eyes. At small $\epsls$ (non-wetting case), the activation energy of viscosity is higher than the one of friction, so that $E_{\text{a,slip}}>0$ and $b$ decreases with $T$. In contrast, at high $\epsls$ (wetting case), the activation energy of friction overcomes the one of viscosity, so that $E_{\text{a,slip}}<0$ and $b$ increases with $T$. A linear regression of this curve gives $E_{\text{a,slip}} \approx -2.94 (\epsls - 0.48)$.
	%The gray vertical dotted lines indicate the melting temperature.
	}
	\label{fig_HT}
\end{figure}

We depict in Fig.~\ref{fig_HT}(a) the slip length of the KA mixture as a function of temperature,
%\hl{for $T$ well above the glass transition temperature of the mixture $T_{\text{g}} = 0.60$} \cite{pedersenPhaseDiagramKobAndersenType2018}
and for different values of the L-S interaction strength $\eps_{LS}$. In this regime, we restrain ourselves to temperatures larger than the glass transition temperature $T_{\text{g}}$, which we estimated to be $T_{\text{g}} \simeq 0.41 \pm 0.01$ using a VFT regression of the viscosity temperature dependence  \cite{vogelViscosity1921,fulcherViscosity1925,tammannViscosity1926} (see below).
%The temperature dependence of the viscosity $\eta$ and the friction coefficient $\lambda$
%are shown in the SM, Fig.~S3. Both 
We find an activation energy  $E_{\text{a}}$ of $2.78$ for the viscosity, and between $1.79$ and $4.39$ for the friction coefficient, depending on $\epsls$.

The slip length $b(T)$ being given by the ratio $\eta / \lambda$, it can be fitted by an Arrhenius law $b(T) \propto \text{exp} (E_{\text{a,slip}}/T)$, with a formal activation energy of slip $E_{\text{a,slip}} = E_{\text{a,viscous}} - E_{\text{a,friction}}$, which can be either positive or negative. Therefore, $b(T)$ can be increasing or decreasing with temperature depending on the relative values of $E_{\text{a,viscous}}$ and  $E_{\text{a,friction}}$. We plot $E_{\text{a,slip}}$ as a function of the L-S interaction energy $\epsls$ in Fig.~\ref{fig_HT}(b). For high values of $\epsls$, $E_{\text{a,friction}}$ becomes larger than $E_{\text{a,viscous}}$. Thus, far from the glass transition temperature, the variation of $b(T)$ is governed by the parameter $\epsls$, which controls the wettability of the system. This is consistent with previous work on LJ liquids \cite{guoTemperatureDependenceVelocity2005}. 

\section{Low temperature regime ($T\leq 1.5$)}

\begin{figure}
	\centering
	\includegraphics[scale=1]{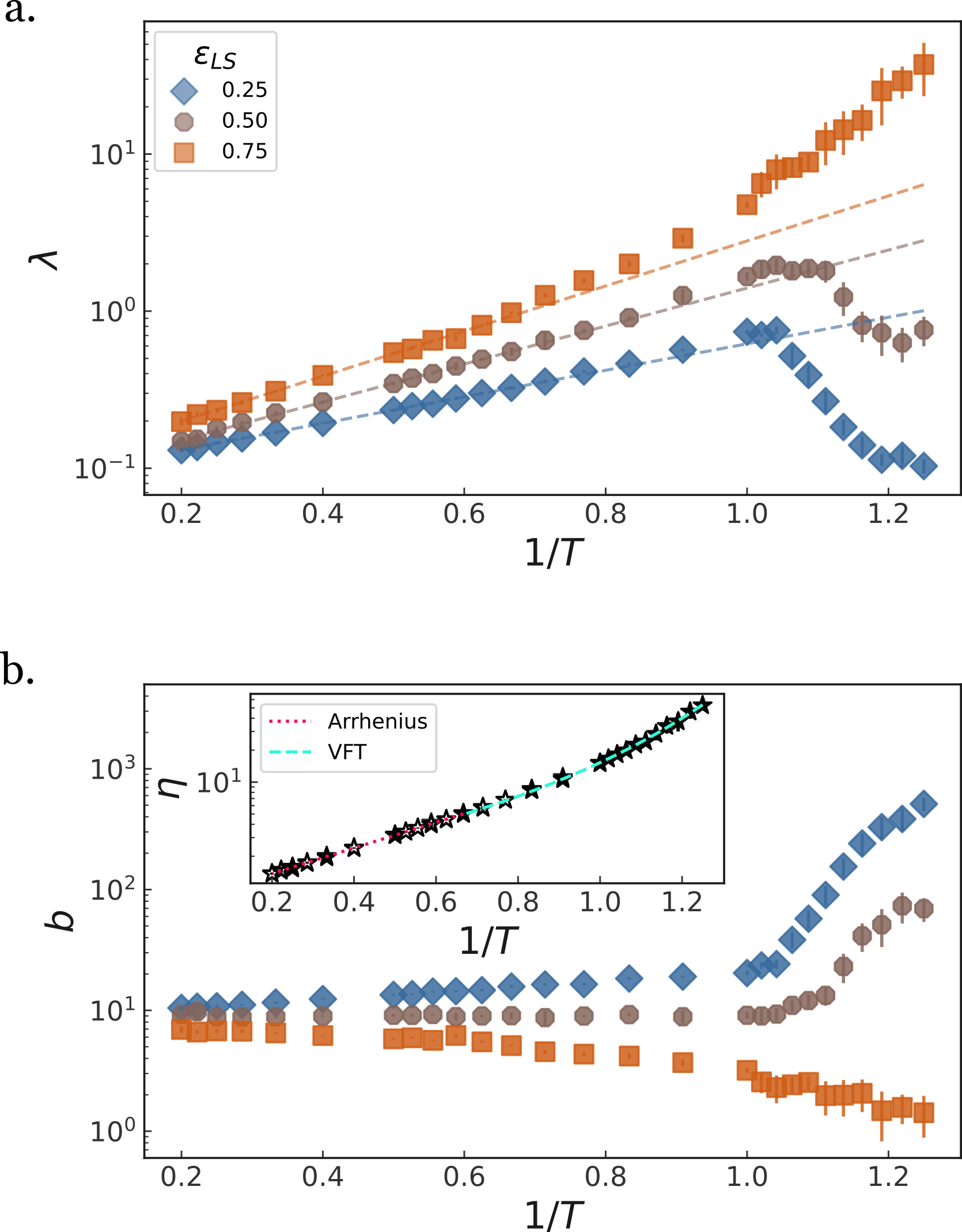}
	\caption{Temperature dependence of (a) the friction coefficient $\lambda(T)$ and (b) the slip length $b(T)$,  derived from $b(T) = \eta(T) / \lambda(T)$. The viscosity $\eta (T)$ is shown in the inset.
	%The gray dotted lines indicate the melting temperature $T_{m}$.
	The dotted lines correspond to Arrhenian regressions at high temperatures. For $\lambda(T)$ and $b(T)$, the colors correspond to different values of $\epsls$ with the same scale as in Fig.~\ref{fig_HT}. For the sake of visibility, only $\epsls = 0.25$ (blue), $0.50$ (brown) and $0.75$ (orange) are shown. For small values of $\epsls$, the friction coefficient drops drastically upon cooling the liquid, so that the slip length $b$ increases sharply.}
	\label{fig_BT}
\end{figure}

We now explore lower temperatures. At these temperatures, the measurement of the friction coefficient $\lambda$ using the Green-Kubo formula becomes delicate because of the so-called plateau problem \cite{espanolForceAutocorrelationFunction1993,bocquetFrictionTensor}. Therefore, in this regime, $\lambda$ is measured with shear simulations only. The viscosity is independent of the value of $\epsls$, as observed in the inset of Fig.~\ref{fig_BT}(b) where the points correspond to measurements at different $\epsls$. At high temperatures, $\eta(T)$ is well described by an Arrhenius law (red curve), while at lower temperatures, $\eta(T)$ can be fitted with a VFT model (blue curve) \cite{vogelViscosity1921,fulcherViscosity1925,tammannViscosity1926}: $\eta = \exp(A + \frac{B}{T-T_{\text{VFT}}})$ with $A = 0.27 \pm 0.06$, $B = 1.45 \pm 0.06$ and $T_{\text{VFT}} = 0.41 \pm 0.01$. For the friction coefficient $\lambda$, we focus on a subset of values for $\epsls$ ($0.25, 0.50$ and $0.75$) for clarity. We observe two different behaviors. For $\epsls = 0.75$, $\lambda$ becomes super-Arrhenian while decreasing the temperature below $1.5$. This slightly overcomes the increase of $\eta$ upon cooling and thus results in a slip length which keeps decreasing while approaching the glass transition. However, for $\epsls = 0.25$ and $0.50$, the friction coefficient suddenly drops by at least one order of magnitude for $T<1$. This corresponds to a strong increase of the slip length at low temperatures by more than one order of magnitude.

\citet{servantieTemperatureDependenceSlip2008} observed the same behavior for the slip length of a LJ polymer slipping on a LJ surface, and attributed it to a difference of mobility between the bulk and the interfacial liquid. In addition, \citet{herreroFastIncreaseNanofluidic2020b} studied the slip length of water on graphene and LJ walls. They observed a moderate increase of the slip length at low temperatures for water on LJ walls, and a strong increase of $b$ for water on graphene, and related them to subtle differences in the temperature evolution of the static and dynamic contributions to viscosity and friction.

To further explore this point, and to understand the fast decrease of $\lambda$ at low temperatures, we have calculated the two-dimensional structure factor $S_\text{liq}(\vec q)$ of the interfacial liquid and compared it to the structure factor $S_\text{wall}(\vec q)$ of the solid wall. The structure factor is calculated by:
\begin{equation}
    S(\vec{q}) = \frac{1}{N} \left[ \left( \sum_{\text{i} = 0}^{N} \cos(\vec{r_{\text{i}}}\cdot\vec{q}) \right)^{2} + \left(\sum_{\text{i} = 0}^{N} \sin(\vec{r_{\text{i}}}\cdot\vec{q}) \right)^{2} \right]
\end{equation}
where $\vec{r_{i}} = x_{i} \vec{e_{x}} + y_{i} \vec{e_{y}}$ is the position of atom $i$ and $N$ is the total number of atoms considered in the calculation (in the first layer of liquid near the wall, delimited by the first non-zero minimum in the density profile in the $z$ direction). The values of $q$ at which we calculate the structure factor are multiples of $2 \pi / L$ with $L$ the size of the box in $x$ and $y$ directions.
The commensurability between the local structure of the liquid interfacial layer and the wall structure is a key factor controlling friction. This commensurability can be quantified by the value of the two-dimensional structure factor of the liquid interfacial layer at the smallest characteristic wavevector of the wall interaction energy landscape, $S_\text{liq}(\vec q_\text{wall})$, where $\vec q_\text{wall}$ is the position of the first peak in the wall structure factor \cite{barratInfluenceWettingProperties1999,falkMolecularOriginFast2010}.

\begin{figure}
	\centering
	\includegraphics[scale=1]{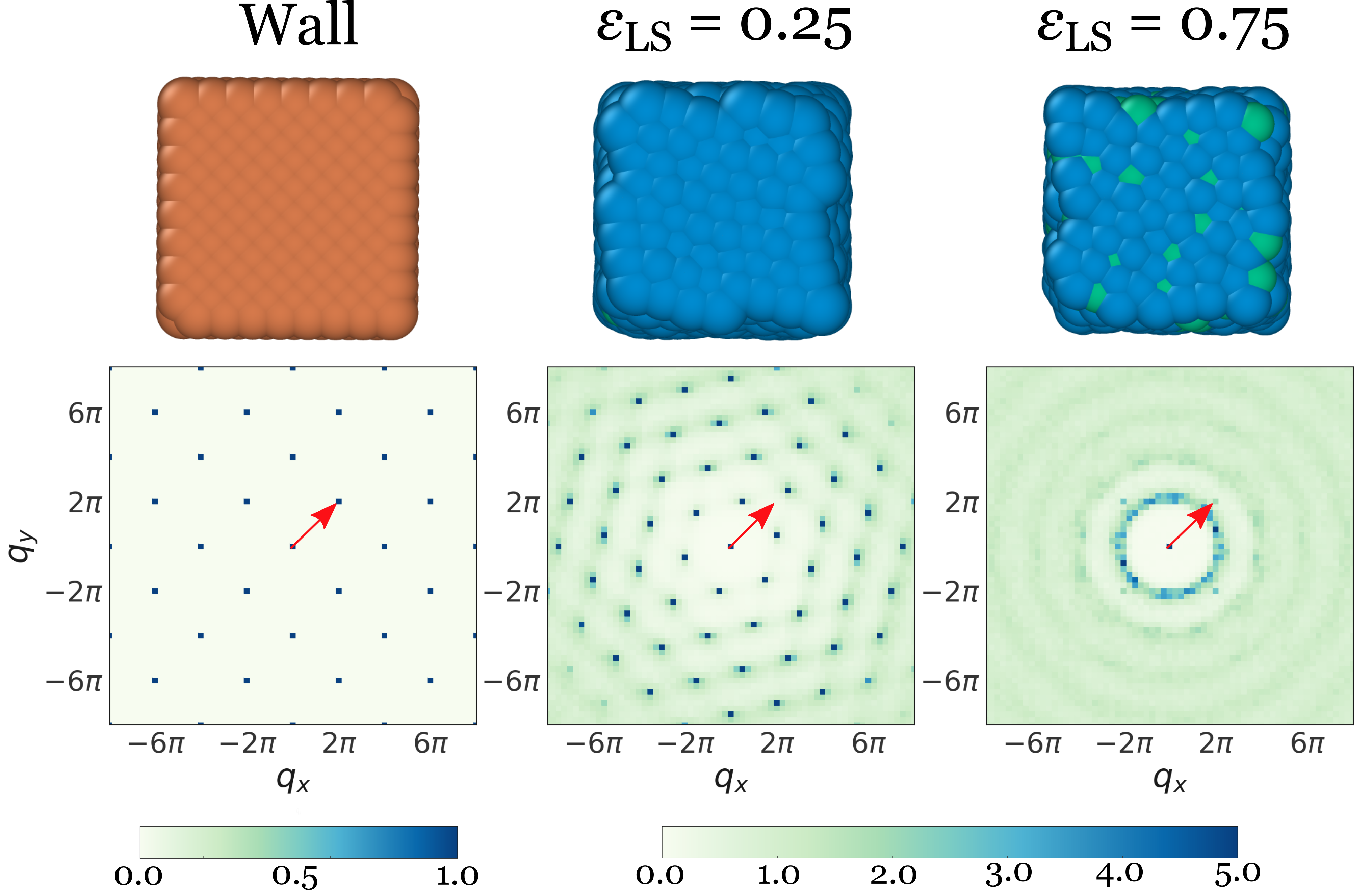}
	\caption{Structure factors of the wall (left) and of the interfacial liquid (middle: $\epsls = 0.25$ and right: $\epsls = 0.75$) at $T = 0.8$, and the corresponding snapshots of the structure (top). The red arrows indicate $\vec{q}_{\text{wall}}$ \textit{i.e.} the vector $\vec{q}$ corresponding to the first peak in the wall structure factor. For small L-S interaction strengths ($\epsls = 0.25$), $B$ particles are expelled from the interface, and the remaining $A$ particles close to the wall structure themselves into a hexagonal lattice, while the wall displays a square structure. Because these two lattices are incommensurate, the friction is highly reduced, leading to very large slip lengths. However, at high L-S interaction strengths ($\epsls = 0.75$), $B$ particles remain at the interface and prevent the near-wall liquid to structure itself, hence the friction remains large and the slip length increases only moderately.}
	\label{fig_Sq}
\end{figure}

The results are shown in Fig.~\ref{fig_Sq}. For $\epsls=0.25$, in the snapshot of the interface, one can see that type B particles are depleted from the interface, allowing the liquid near the wall to structure itself into a lattice, which turns out to be hexagonal. We have quantified this depletion of B particles near the wall in Fig.~S7 of the SM where we plot the concentration of A particles near the wall against temperature. For low values of $\epsls$ at low temperatures, the concentration of A particles at the interface is close to $100\%$ which allows the corresponding liquid layers to crystallise. Because the wall displays an incommensurate square lattice, the liquid structure factor is very small at the position $\vec q_\text{wall}$ of the first peak of the wall structure factor, i.e., $S(\vec q_\text{wall}) \ll 1$, and the friction coefficient $\lambda$ is strongly reduced.
It is worth noting that the lattice of the interfacial liquid displays an hexagonal order both at equilibrium and under shear, at any considered velocity (see the SM, Fig.~S6). In contrast, for stronger L-S interaction ($\epsls=0.75$), type B particles remain at the interface and prevent the interfacial liquid to structure itself, and thus its structure factor remains that of a liquid. In this case, $S(\vec q_\text{wall})$ remains on the order of $1$, and the friction coefficient still increases in an Arrhenian way upon cooling down the liquid, which results in a moderate increase of the slip length.

Here, superlubricity is possible because of the structure of the interface, and is reminiscent of solid-solid superlubricity, as evidenced experimentally e.g. for graphite \cite{dienwiebel2004superlubricity}. Indeed, the role of incommensurability in reducing the friction between two solids has been reported by Zhang \textit{et al.} \cite{zhangAtomicSimulationsKinetic2005} and Franchini \textit{et al.} \cite{franchiniEffectsCommensurabilityDisorder2011} for Al/Al and Xe/Cu interfaces respectively. In addition, Cieplak \textit{et al.} \cite{cieplakMolecularOriginsFriction1994} have observed a strong reduction of friction from a fluid to a crystallized layer of krypton adsorbed on gold because of incommensurability between the crystallized krypton layer and the gold lattice.

\section{Conclusion}

In conclusion, the MD investigation of $b(T)$ has revealed that at high temperatures, not only the viscosity but also the friction coefficient follow an Arrhenius law. Therefore, the slip length may be also described by an Arrhenius law with an effective activation energy that can be either positive or negative depending on the strength of the liquid-solid interaction potential $\epsls$. This result aligns well with the prediction for polymer melts described in \cite{henotTemperatureControlledSlipPolymer2018}. 

A more striking phenomenon is that, at lower temperatures, the slip length may increase by orders of magnitude. This massive enhancement is the result of both the super-Arrhenian temperature dependence of the viscosity in the supercooled regime and the sub-Arrhenian behavior of friction with cooling. In particular, for weak L-S interactions, the friction coefficient is highly reduced due to the emergence of a local fluid structure, which is incommensurate with the solid lattice. These conditions are highly favorable to observe giant values of the slip length. For stronger L-S interactions, the friction coefficient increases in a super-Arrhenian way, resulting in merely a moderate decrease of the slip length. 
These results showing a possible mechanism for massive slippage on low energy surfaces could lead to promising transport applications in nanofluidics, and call for experiments probing the slip length at supercooled liquid-solid interface. Short polymer melts close to their glass transition temperature flowing over weakly interacting surfaces are good candidates to evidence massive temperature dependent slip lengths.

\begin{acknowledgments}
The authors thank Cecilia Herrero, Etienne Fayen and Patrick Judenstein for fruitful discussions. 
We are also grateful for HPC resources
from GENCI/TGCC (grant A0090810637), from the PSMN mesocenter in Lyon and from the THEO group in the LPS, Orsay. We thank the ANR POILLU (grant ANR-19-CE06-007) as well. 
\end{acknowledgments}

\bibliography{biblio_all}

\end{document}

%\preprint{APS/123-QED}

\title{Supplemental material for: ``Giant slip length at a supercooled liquid-solid interface''}

\author{Suzanne Lafon}
 \email{suzanne.lafon@universite-paris-saclay.fr}
 \affiliation{Paris-Saclay University, CNRS, Solid State Physics Laboratory, 91405 Orsay, France}
\author{Alexis Chennevière}
 \affiliation{CEA Saclay, Léon Brillouin Laboratory, 91191 Gif-sur-Yvette, France}
\author{Frédéric Restagno}
 \affiliation{Paris-Saclay University, CNRS, Solid State Physics Laboratory, 91405 Orsay, France}
\author{Samy Merabia}
 \affiliation{\ilm}
\author{Laurent Joly}
 \affiliation{\ilm}

\date{\today}

\maketitle

\tableofcontents

\section{Measurement of viscosity and friction coefficient}

\subsection{Measurement under shear}

When a liquid is sheared by two walls, its viscosity is determined through $\eta = \tau_{\text{LS}}/\dot{\gamma}$, with $\tau_{LS}$ being the stress exerted by the liquid on the walls and $\dot{\gamma}$ being the shear rate extracted from the velocity profile. Then, the friction coefficient is measured by $\lambda = \tau_{\text{LS}}/\Vs$, with the slip velocity $\Vs = U - \dot{\gamma} h/2$ where $h$ is the hydrodynamic height of the system. 
%wall position (HWP). 
\citet{herreroShearForceMeasurement2019} have shown that $h$ can be computed by identifying the hydrodynamic wall position with the Gibbs dividing plane (GDP).
The way we determine the GDP is illustrated in Fig.~\ref{SI_GDP}. Near the walls, the density profile $n(z)$ oscillates. The number of particles between the wall and a given height $z_{0}$ inside the bulk is calculated by integrating the liquid density $n(z)$ between $0$ and $z_{0}$: $N = \int_{0}^{z_{0}} n(z)\dd z$. A homogeneous distribution of these $N$ atoms would lead to $N = n_{z}^\text{bulk} (z_{0}-z_{s})$ with $z_{s}$ the GDP position. By equating these two equations, one can calculate $z_{s}$:
\begin{equation}
    z_{s} = z_{0} - \frac{\int_{0}^{z_{0}} n(z)\dd z}{n_{z}^\text{bulk}} . 
\end{equation}

\begin{figure}[!ht]
	\centering
	\includesvg[scale=1]{Figure_S1.svg}
	\caption{Density profile of the liquid near the bottom wall. Close to the wall, the density profile displays oscillations. The dots represent measured values of the density, while the light green area represents the integral of the density between $0$ and some reference height in the bulk $z_{0}$. The dark green area represents a homogeneous liquid containing the same number of atoms as the real liquid. The GDP $z_{s}$ is determined by equating the light green and the dark green areas.}
	\label{SI_GDP}
\end{figure}

\begin{figure}[!ht]
	\centering
	\includesvg[scale=1]{Figure_S2.svg}
	\caption{Threshold shear velocity $U_{\text{th},\lambda}$ as a function of temperature for two different values of the liquid-solid interaction strength $\epsilon_{\text{LS}}$. Both the liquid viscosity $\eta$ and the friction coefficient $\lambda$ are independent of the shear-velocity $U$ for $U \leq U_{\text{th},\lambda}$.}
	\label{SI_Uth}
\end{figure}

At sufficiently low shear rate $\dot{\gamma}$, the slip length is constant, while at high shear rates, the Navier condition fails and the slip length increases rapidly with $\dot{\gamma}$. We want to stay in the linear response regime so, for each system, we apply various shear velocities $U$ to determine the threshold velocity above which we leave the Navier regime. In practice, we measure the viscosity $\eta$ and the friction coefficient $\lambda$. At low $U$, we have a plateau while at high $U$, both $\eta$ and $\lambda$ decrease. We computed the value of the threshold shear velocity for both $\eta$ and $\lambda$. The threshold velocity was smaller for $\lambda$ than for $\eta$, therefore we only kept this one as an upper bound for the shear velocity. In Fig.~\ref{SI_Uth}, we plot the threshold velocity for $\lambda$ and the corresponding threshold shear rate as a function of temperature and L-S interaction strength.

We do at least $3$ different measurements for each data point shown in the results. At low temperatures ($T \leq 1$), the threshold velocity becomes very small so the curves become noisy and therefore we do at least $6$ different measurements for each data point.

\subsection{Measuring procedure}

\begin{figure}[!ht]
	\centering
	\includesvg[scale=1.0]{Figure_S3.svg}
	\caption{Left: velocity profiles for one given measurement at $U = 0.10$, $T = 0.90$ and $\epsilon_{\text{LS}} = 0.25$. The colors correspond to different times during shear. Middle: shear rate $\dot{\gamma}$ for different measurements performed in the same conditions. The velocity profiles plotted on the left correspond to measurement $4$ in this graph. Right: Shear rate as a function of $1/T$, measured in the newtonian regime at $\epsilon_{\text{LS}} = 0.25$.}
	\label{SI_vxinstable}
\end{figure}

For a given system, we plot the velocity profile (see Fig.\ref{SI_vxinstable}, left). At low temperatures, the velocity profile evolves with time. We attribute this time dependency to the consequence of a very high slip length, which allows the liquid to diffuse in block, adding a random constant velocity to the shear flow. However, the slope $\dot{\gamma} = \frac{\partial v}{\partial z}$ is approximately constant over time. We measure the shear rate at different times and for systems having identical parameters but different initial conditions (see Fig. \ref{SI_vxinstable}, middle). Then the shear rate at a given $(T,\epsilon_{\text{LS}})$ is taken as the mean value of the shear rates at all times for all measurements done in the linear response regime. The error bars correspond to the standard deviation of these measurements divided by the square-root of the number of measurements.

\subsection{Effect of pressure on slip length}

For all the simulations, we have chosen a pressure $P = 10.0$. This is a standard choice for a Kob-Andersen liquid. We have tested the effect of pressure on the slip length for pressures around our reference $P = 10.0$. The results are shown in Fig.\ref{SI_effet-P}. Both the viscosity and the friction coefficient increase rather linearly with the pressure. The slip length - which is the ratio of the two - appears to remain relatively constant in the range of explored pressures.

\begin{figure}[!ht]
	\centering
	\includesvg[scale=1.2]{Figure_S4.svg}
	\caption{Effect of pressure on the slip length. The measurements have been done for a sheared liquid at $U = 0.40$ and $T = 2.0$.}
	\label{SI_effet-P}
\end{figure}

\subsection{Effect of wall density on slip length}

We have studied the evolution of the slip length as a function of the wall particles spacing $a$ which is related to the reduced wall density $\tilde{\rho}$ through $\tilde{\rho} = 4 \sigma^{3}/a^{3}$. The result is shown in Fig.~\ref{SI_effet-rho}. We can see that the slip length for $a = 1.0$ (which corresponds to $\tilde{\rho} = 4.0$) is ten times higher than the one for $a \approx 1.6$ (which corresponds to $\tilde{\rho} = 1.0$).
%\hl{Therefore, we call the first one "slipping wall" and the second one "non-slipping wall".}

\begin{figure}[!ht]
	\centering
	\includesvg[scale=1.0]{Figure_S5.svg}
	\caption{Effect of wall particles spacing on the slip length. The measurements have been done for $\epsls = 0.25$ using Green-Kubo simulations.}
	\label{SI_effet-rho}
\end{figure}

\section{Structuration at the liquid-solid interface at low $T$}

\subsection{Structure factors}

\begin{figure}[!ht]
	\centering
	\includegraphics[scale=0.8]{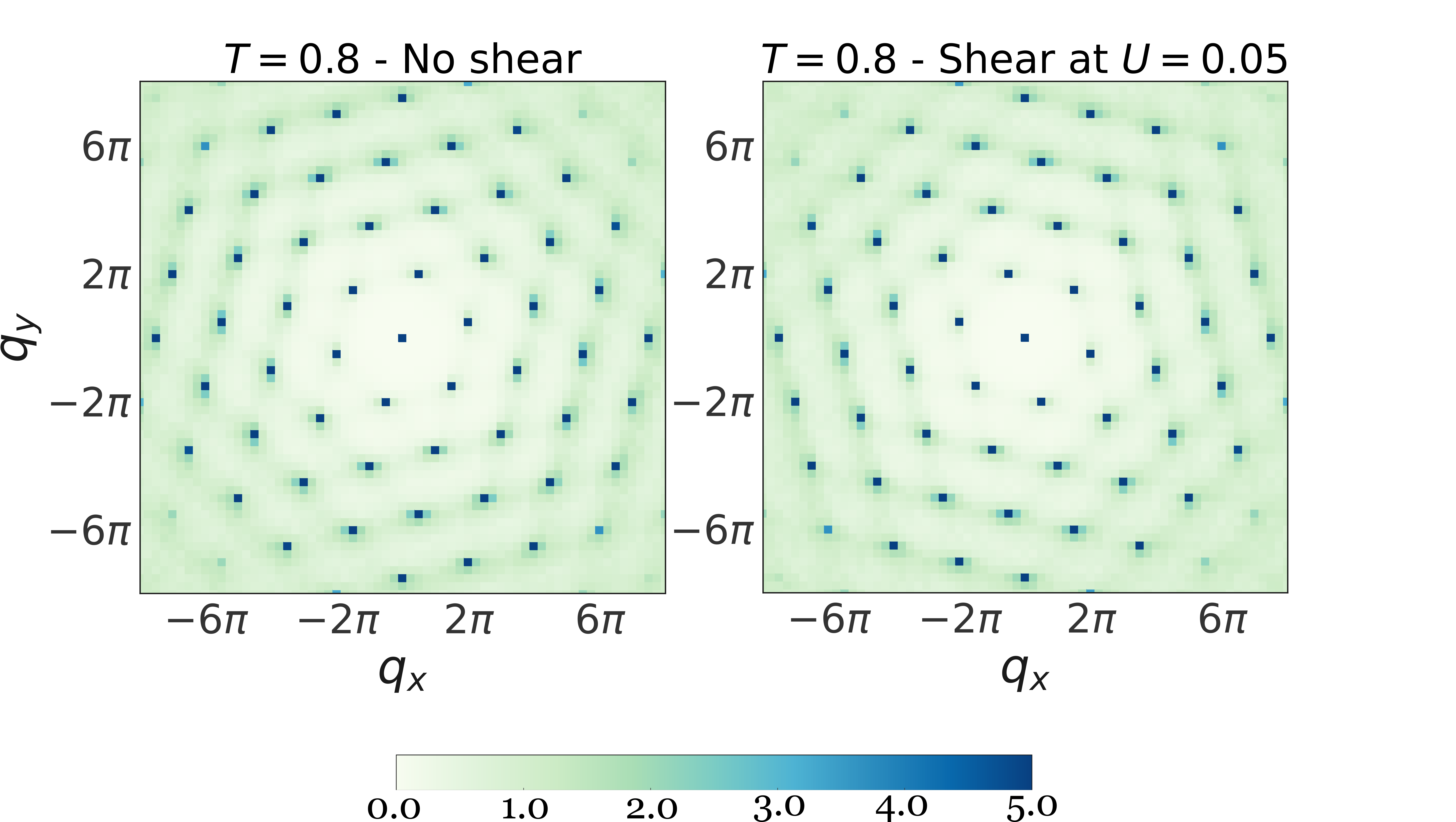}
	\caption{Comparison of the structure factor of the interfacial liquid. Left: $T = 0.8$  without shearing the liquid. Right: with shearing the liquid at $U = 0.05$. The structure of the liquid layer in contact with the wall remains hexagonal both at rest and under shear. The only difference is that the lattice can be rotated.}
	\label{SI_Sq}
\end{figure}

We calculate the structure factor of the first liquid layer in contact with the solid wall, for systems under shear or at rest, and for various temperatures. We observe that a hexagonal lattice appears at low temperatures for weak liquid-solid interaction strengths. The structure of the liquid layer remains hexagonal even under shear (see Fig.~\ref{SI_Sq}).

\subsection{Discussion about demixing}

\begin{figure}[!ht]
	\centering
	\includegraphics[scale=1.0]{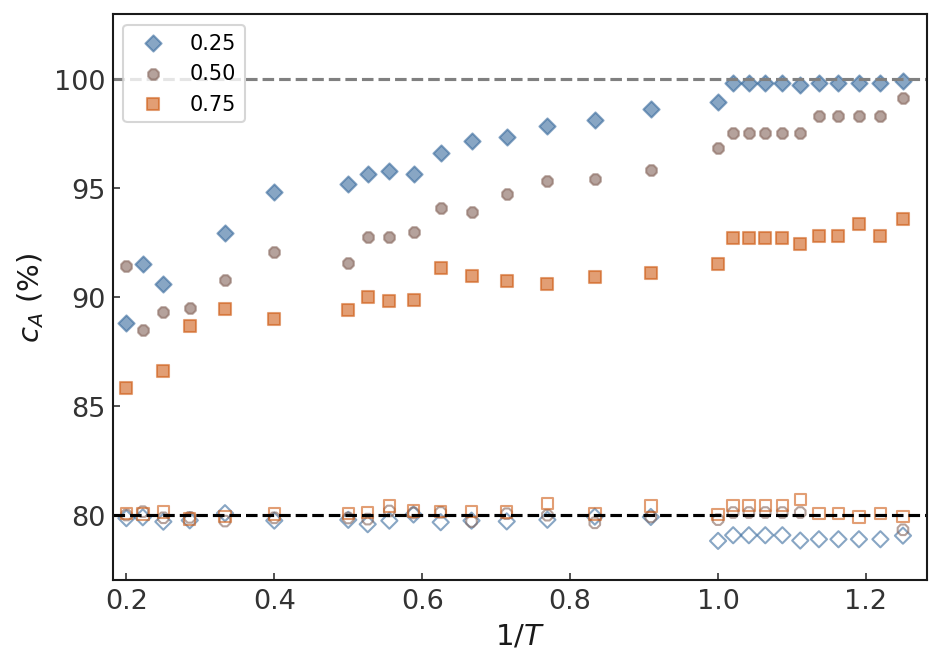}
	\caption{Concentration of $A$ particles close to the interface, as a function of temperature and for different values of $\epsilon_{\text{LS}}$. Empty symbols correspond to concentration of $A$ particles in the bulk.}
	\label{SI_demixion}
\end{figure}

In order to explain the crystallisation near the wall at low temperatures, we have computed the concentration of A particles $c_{\text{A}}$ near the wall as a function of temperature. For each measurement, we compute the position of the first liquid atoms $z_{min}$ near the wall and we define a layer of thickness $0.1$ inside which we calculate the concentrations. $c_{\text{A}}$ is defined as $c_{\text{A}} = N^{interface}_{A}/N^{interface}$ where $N^{interface}_{A}$ ($N^{interface}$) is the number of A particles (the total number of particles) in this first liquid layer. The result is shown in Fig.~\ref{SI_demixion}. We observe that $c_{\text{A}}$ near the interface is always larger than in the bulk. In addition, as the temperature decreases, $c_{\text{A}}$ increases and even reaches $100\%$ at $T\leq 1$ for $\epsls = 0.25$. The ability of the liquid to be supercooled is due to its bi-dispersity. Therefore, when the liquid becomes almost pure, it cannot maintain its supercooled state and crystallises.

\bibliography{biblio_single}